\documentclass[aps,prx,twocolumn,groupedaddress,floatfix,longbibliography]{revtex4-2}

\usepackage{graphicx}
\usepackage{dcolumn}
\usepackage{bm}
\usepackage{amsmath}
\usepackage{xcolor}
\usepackage{braket}
\usepackage[utf8]{inputenc}
\usepackage[T1]{fontenc}
\usepackage{hyperref}
\hypersetup{
breaklinks=true,
    colorlinks=true,
    linkcolor=blue,
    citecolor=blue,
    filecolor=magenta,
    urlcolor=cyan,
}

\usepackage{blkarray}
\usepackage{multirow}

\usepackage{natbib}      

\newcommand{\cc}{{\cal C}}

\newcommand{\bG}{{\bf G}}
\newcommand{\br}{{\bf r}}

\begin{document}

\title{Wigner-molecule supercrystal in transition-metal dichalcogenide moir\'e superlattices:
Lessons from the bottom-up approach}
\author{Constantine Yannouleas}
\email{Constantine.Yannouleas@physics.gatech.edu}
\author{Uzi Landman}
\email{Uzi.Landman@physics.gatech.edu}

\affiliation{School of Physics, Georgia Institute of Technology,
             Atlanta, Georgia 30332-0430}

\date{04 December 2023}

\begin{abstract}
The few-body problem for $N=4$ fermionic charge carriers in a double-well moir\'{e} quantum dot
(MQD), representing the first step in a bottom-up strategy to investigate formation of molecular
supercrystals in transition metal dichalcogenide (TMD) moir\'e superlattices with integral fillings,
$\nu > 1$, is solved exactly by employing large-scale  
exact-diagonalization via full configuration interaction (FCI) computations. 
A comparative analysis with the mean-field solutions of the often used spin-and-space unrestricted Hartree Fock (sS-UHF) demonstrates the limitations of the UHF method
(by itself) to provide a proper description of the influence of the interdot Coulomb interaction. In
particular, it is explicitly shown for $\nu=2$ that the exact charge densities (CDs) within each MQD 
retain the ring-like shape characteristic (for a wide range of relevant parameters) of a fully 
isolated MQD, as was found for {\it sliding\/} Wigner molecules (WMs). 
This deeply quantum-mechanical behavior contrasts sharply with the UHF CDs that portray solely 
orientationally pinned and well localized dumbbell dimers. An improved CD, which agrees with the 
FCI-calculated one, derived from the restoration 
of the sS-UHF broken parity symmetries is further introduced, suggesting a beyond-mean-field 
methodological roadmap for correcting the sS-UHF results.
It is conjectured that the conclusions for the $\nu=2$ moir\'e TMD superlattice case extend to all 
cases with integral fillings that are associated with sliding WMs in isolated MQDs.
The case of $\nu=3$, associated with a pinned WM in isolated MQDs, is an exception.  
\end{abstract}   

\maketitle

\begin{figure}[t]
\centering\includegraphics[width=8.0cm]{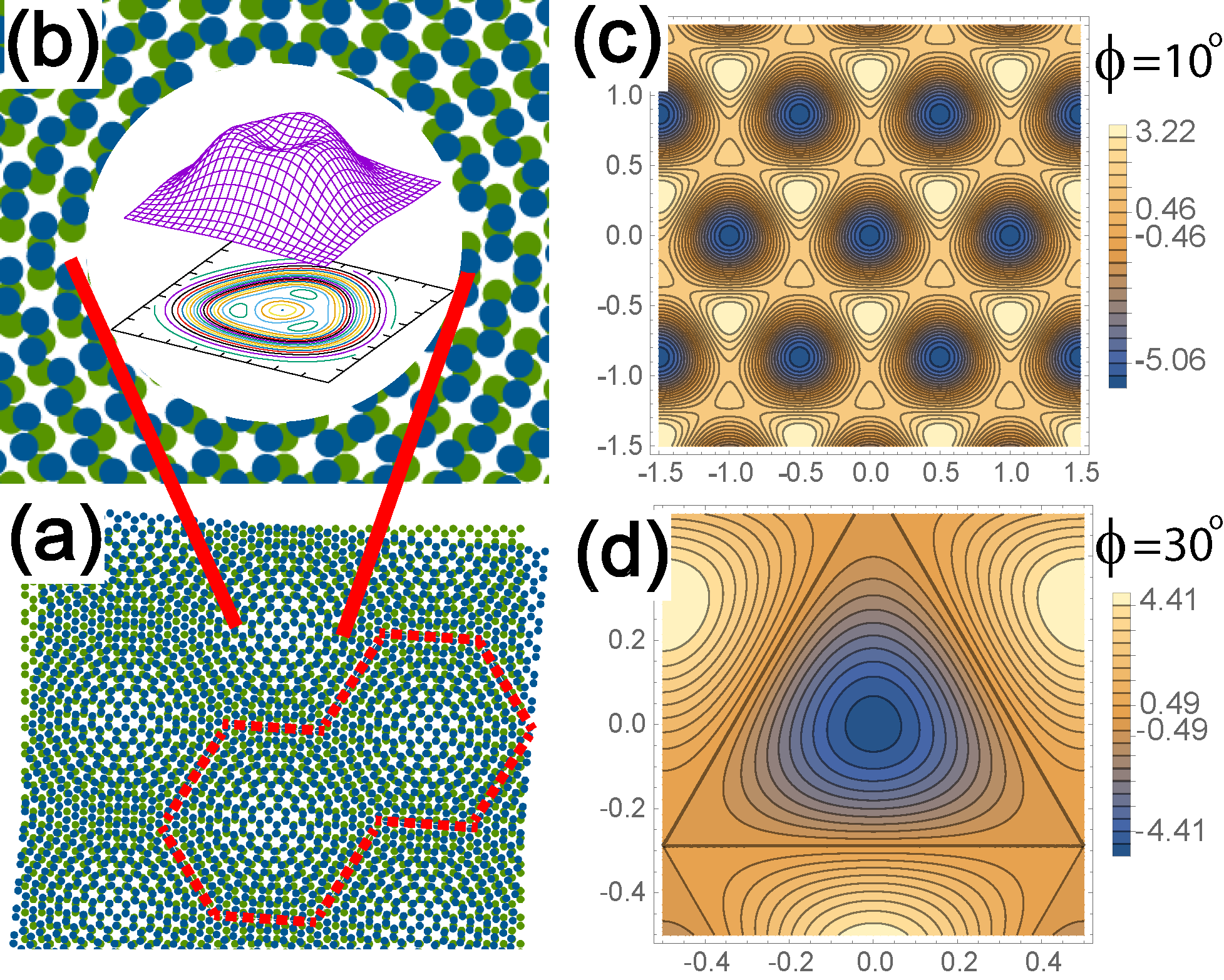}
\caption{
(a) Schematic moir\'e pattern produced by two twisted monolayers. The area demarcated by the
dashed red line corresponds to the isolated moir\'e {\it double\/} quantum dot investigated in this
paper. 
(b) The spin-singlet ground-state FCI charge density associated with $N=2$ holes in a 
{\it single\/} isolated moir\'e
quantum dot, exhibiting a ring-like shape with a trilobal distortion. This ring-like CD is 
characteristic of a {\it sliding\/} (contrasted to a {\it pinned\/}) Wigner molecule.
Parameters used: effective mass $m^*=0.9m_e$, dielectric constant $\kappa=5$, strength of
moir\'e modulation $v_0=10.3$ meV, moir\'e lattice constant $a_M=9.8$ nm, trilobal distortion 
$\phi=20^\circ$; see Eq.\ (\ref{mpot}). These parameters are also used in all CD calculations 
(either FCI or UHF) for the MDQD case below. 
(c) Broader view of the moir\'e periodic potential structure given by Eq.\ (\ref{mpot}) for an angle 
of $\phi=10^\circ$.
(d) Potential of a single moiré QD for $\phi= 30^\circ$.
In (c) and (d), $v_0=15$ meV and $a_M=14$ nm.
Length units are in nm in (b) and in $a_M$ in (c) and (d).
CD in (a) in units of 1/nm$^2$. All CDs in this paper are normalized to $N=4$.  
}
\label{pots}
\end{figure}

\begin{figure}[t]
\centering\includegraphics[width=7.5cm]{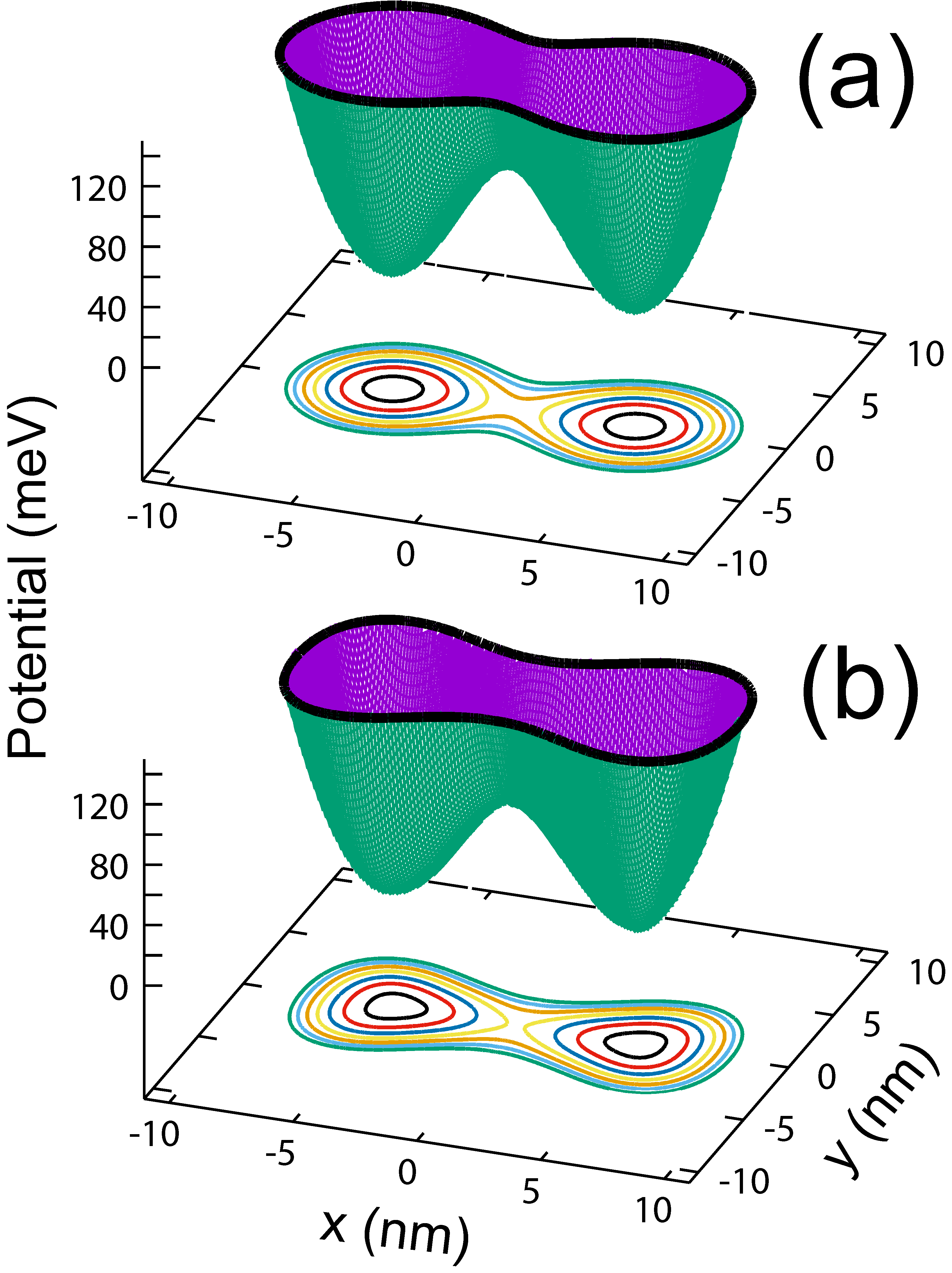}
\caption{
The confining potential for the MDQD. (a) The TCO potential according to Eq.\ (\ref{vtco}), which
does not include the trilobal deformation. (b) The potential [Eq.\ (\ref{vmdqd})] for the MDQD,
which does include the trilobal deformation within each QD. Parameters used: $\hbar \omega_0=
36.71$ meV (which corresponds to $v_0=10.3$ meV, $a_M=9.8$ nm, and $\phi=20^\circ$), $-x_1=x_2=4.9$
nm, $\epsilon^b=0.44$, and $f=0.15$. The interdot barrier is 84.07 meV in (a) and 71.46 meV in (b). 
}
\label{potsdd}
\end{figure}

Understanding the electronic spectral and configurational organization beyond that of natural 
atoms is rapidly becoming a major research direction focusing on the exploration of the nature of 
a few charged carriers trapped in artificially fabricated, isolated or superlattice-assembled, 
quantum dots (QDs) \cite{kouw97,hans07,jing22}. Such research is motivated by the potential for
utilizing these systems, with high tunability and control, in future quantum information and 
computational platforms \cite{copp13,vand19,deng20,burk23}. Earlier studies have unveiled a novel 
fundamental-physics aspect in such nanosystems, namely formation of quantum Wigner molecules 
(WMs), originally predicted theoretically \cite{yann99,grab99,yann00,fili01,yann02.2,mikh02,
harj02,yann03,szaf03,yann04,szaf04,roma06,yann06.3,yann07,yang07,yann07.2,yann07.3,umri07,yang08,
roma09,yann15,yann21,erca21.2,urie21,yann22,yann22.2} in two-dimensional (2D) semiconductor QDs, 
as well as trapped ultracold atoms, and subsequently observed experimentally in GaAs QDs 
\cite{yann06,kall08,kim21,kim23}, Si/SiGe QDs \cite{corr21}, and carbon-nanotubes \cite{peck13}. 
Remarkably, recent work \cite{yann23} extended the WM portfolio to the newly arising and highly 
pursued field of TMD moir\'e materials, owing mainly to the promise for fundamental-physics 
discoveries and the potential for advancing quantum-device applications. 

Adopting a bottom-up methodology, and building on the demonstrated emergence \cite{yann23} of WMs in
the quasi-isolated moir\'e pockets [most often referred to as moir\'e quantum dots (MQDs)], we 
address here the inevitable incorporation of such single MQDs in a superlattice structure. 
Specifically, this paper focuses on the effects on WM formation
resulting from the interaction between neighboring MQDs. Two different methodologies will be used
in this endeavor, namely: (i) the spin-and-space unrestricted Hartree-Fock (sS-UHF) 
\cite{yann99,yann02,yann02.2,yann07,yann00.2}
and (ii) the full configuration interaction (FCI) \cite{shav98,yann03,szaf03,yann06.2,ront06,yann07,
yann07.3,yann15,yann21,yann22.2,yann22.2,yann22.3,szabo}.
In particular, through a detailed comparison with the FCI {\it exact\/} resuts (serving here as 
comparative benchmarks), we show that except when the classical equilibrium configuration of the 
$\tilde{n}$ confined carriers in each single MQD is commensurate with the trilobal symmetry 
of the moir\'e confinement (see below, leading to a pinned WM 
configuration, e.g., when $\tilde{n}=3$), the sS-UHF approximation (by itself) is unreliable for 
investigations of WM-exhibiting moir\'e double quantum dots (MDQDs) in {\it unstrained\/} TMD 
bilayers (that is in naturally occurring, bias-free, cases). This 
deficiency dictates further corrective measures (beyond
the mean-field sS-UHF) that are provided by the theory of restoration of broken symmetries and 
extensions thereof \cite{yann02.2,yann02,yann04.2,yann06.2,yann06.3,roma06,yann07,shei21,ringbook}.
Specifically, for $\nu=2$ and for a set of materials parameters suitable to moir\'e TMD 
superlattices, we show that, in spite of the interaction with a neighboring MDQ, the ground-state 
FCI charge densities within each MQD remain ring-like with a superimposed trilobal distortion, as in 
the case of a single isolated MQD \cite{yann23} illustrated in the inset of Fig.\ \ref{pots}(a). This 
contrasts sharply with the corresponding (mean-field) sS-UHF CD {\it prior\/} to symmetry 
restoration, which exhibits a pair of two antipodal and well localized charge carriers.

On the other hand, the {\it symmetry-restored\/} UHF (SR-UHF) charge densities, are in agreement with 
the exact (FCI) ones. These results suggest a much desired gateway for systematic large-scale 
computational studies of WM-MQD assemblies in TMD materials using a beyond-mean-field-corrected 
SR-sS-UHF methodology, capable of modeling systems comprised of a much larger number of carriers 
(electrons or holes) that may be treated with the exact-diagonalization, FCI, method. Furthermore, in
light of rapid advances in STM imaging techniques \cite{wang21,yazd23}, we expect that the FCI 
predictions shown here will gain verification in the near future (see added note at the end).

\begin{figure}[t]
\centering\includegraphics[width=8.0cm]{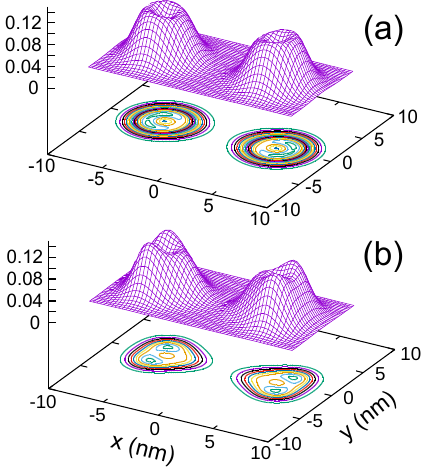}
\caption{
Ground-state FCI charge densities for the case of the moir\'e double QD with $N=4$ holes. 
(a) CD for the confinement in Fig.\ \ref{potsdd}(a) (no trilobal deformation). 
(b) CD for the confinement in Fig.\ 
\ref{potsdd}(b) (trilobal deformation included). Remaining parameters: effective mass $m^*=0.90m_e$, 
dielectric constant $\kappa=5$. See text for a detailed description. CDs in units of 1/nm$^2$.
}
\label{fcicd}
\end{figure}

\begin{figure}[t]
\centering\includegraphics[width=8.0cm]{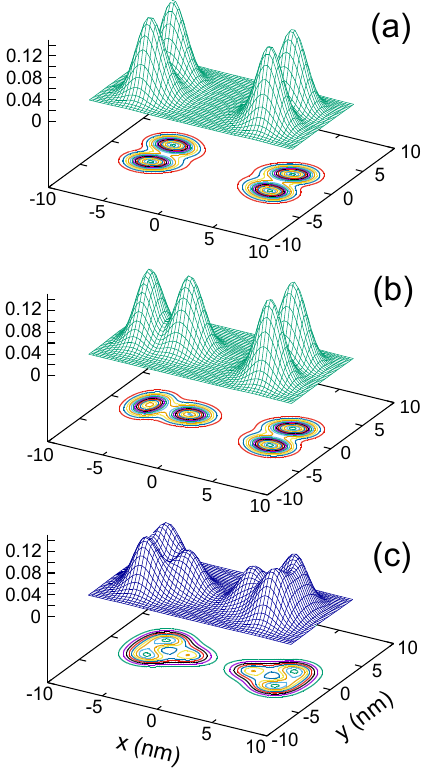}
\caption{
(a) and (b) Charge densities for $N=4$ holes associated with the two lowest-energy UHF isomers 
(with $S_z=0$) for the
MDQD confinement displayed in Fig.\ \ref{potsdd}(b) (which includes the trilobal deformation within 
each MQD). In both cases, the UHF CDs consist of dumbbell-like pairs of well localized charge 
carriers. In (a), both the left and right pairs have axes oriented perpendicular to the $x$-axis. In 
(b), the axis of the left dumbbell is parallel to the $x$ axis, whilst the right dumbbell remains 
perpendicular to the $x$-axis.    
(c) CD of the $x$-parity restored wave function associated with the UHF isomer in panel (b).
Note that, unlike the pure UHF CDs in (a) and (b), the $x$-parity-restored CD in (c) exhibits
a ring-like shape in good agreement with the FCI (exact) CD in Fig.\ \ref{fcicd}(b). See text for a 
detailed description. Effective mass $m^*=0.90m_e$ and dielectric constant $\kappa=5$.
CDs in units of 1/nm$^2$.
}
\label{uhfcd}
\end{figure}

{\it Confinement potentials and many-body Hamiltonian.\/}
The potential confining the extra charge carriers at the pockets of the 2D moir\'{e} superlattice
can be approximated by the expression \cite{macd18,ange21,fu20}
\begin{align}
V(\br) = -2 v_0 \sum_{i=1}^3 \cos(\bG_i \cdot \br + \phi),
\label{mpot}
\end{align}
where $\bG_i=[ (4\pi/\sqrt{3}a_M) ( \sin(2\pi i/3), \cos(2\pi i/3) ) ]$ are the moir\'{e} reciprocal
lattice vectors. The materials specific parameters of $V(\br)$ are $v_0$ (which can also be 
experimentally controlled through voltage biasing), the moir\'{e} lattice constant $a_M$, and the 
angle $\phi$. $a_M$ is typically of the order of 10 nm, which is much larger than the lattice 
constant of the monolayer TMD material (typically a few \AA). For the overall periodic-array 
structure of $V(\br)$, see Fig.\ \ref{pots}(c). 

The parameter $\phi$ controls the strength of the trilobal $C_3$ crystal-field-type anisotropy in 
each MQD potential pocket; see Fig.\ \ref{pots}(d). This anisotropy can be seen by expanding 
$V(\br)$ in Eq.\ (\ref{mpot}) in powers of $r$, and defining an approximate confining potential, 
$V_{\rm MQD}(\br)$, for a single MQD as follows:
\begin{align}
\begin{split}
V_{\rm MQD}(\br) \equiv V(\br) + 6 v_0 \cos(\phi)  \approx m^* \omega_0^2 r^2/2 + 
\cc \sin(3 \theta) r^3.
\end{split}
\label{vexp}
\end{align}
with $m^*\omega_0^2=16 \pi^2 v_0 \cos(\phi)/a_M^2$, and 
$\cc=16 \pi^3 v_0 \sin(\phi)/( 3\sqrt{3}a_M^3)$; $m^*$ is the effective mass and the expansion of 
$V(\br)$ can be restricted to the terms up to $r^3$. $(r, \theta)$ are the polar coordinates of
the position vector $\br$.

We construct a potential confinement for an isolated {\it pair\/} of two neighboring MQDs [see the 
area marked by the thick dashed red border in Fig.\ \ref{pots}(a)] in two steps: 

First, we consider the potential of a two-center-oscillator (TCO) with a smooth neck 
\cite{yann99,yann07,yann09,yann22,yann22.2,yann22.3}. Namely, 
\begin{equation}
V_{\rm TCO}(x,y) = \frac{1}{2} m^* \omega^2_y y^2
    + \frac{1}{2} m^* \omega^2_{x k} x^{\prime 2}_k + V_{\rm neck}(x),
\label{vtco}
\end{equation}
where $x_k^\prime=x-x_k$ with $k=1$ for $x<0$ (left) and $k=2$ for $x>0$ (right). $y$ denotes the 
coordinate perpendicular to the interdot axis ($x$). In this paper, we take 
$\omega_{x1}=\omega_{x2}=\omega_y=\omega_0$, with $\omega_0$ coinciding with that of a single
MQD [see Eq.\ (\ref{vexp})].

For the smooth neck, we use
\begin{align}
V_{\rm neck}(x) = \frac{1}{2} m^* \omega_0^2
\Big[ {\cal C}_k x^{\prime 3}_k + {\cal D}_k x^{\prime 4}_k \Big] \Theta(|x|-|x_k|),
\label{vneck}
\end{align}
where $\Theta(u)=0$ for $u>0$ and $\Theta(u)=1$ for $u<0$. The four constants ${\cal C}_k$ and 
${\cal D}_k$ can be expressed via two parameters, as follows: ${\cal C}_k= (2-4\epsilon^b)/x_k$ and
${\cal D}_k=(1-3\epsilon^b)/x_k^2$, where the barrier-control parameter $\epsilon^b=V_{b}/V_{0}$ is
related to the height of the targeted interdot barrier $V_{b}$, and $V_{0}=m^*\omega_0^2 x_k^2/2$.
The $V_{\rm TCO}$ potential is illustrated in Fig.\ \ref{potsdd}(a).

Second, we intoduce the trilobal deformation in each MQD through the expression
\begin{align}
V_{\rm MDQD}(x,y)=V_{\rm TCO}(x,y) \big( 1+f \sin[ 3\theta^\prime_k + (-1)^k \pi/2 ] \big),
\label{vmdqd}
\end{align}
where $\theta^\prime_k$ is the counterclockwise angle around the point $(x_k^\prime,0)$, with
$x_k^\prime$ defined as in Eq.\ (\ref{vtco}).
The factor $f$ is taken such that the modified interdot barrier $V_b(1-f)$ equals
the minimum barrier 
between the two MQDs, as determined by the original moir\'e potential in Eq.\ (\ref{mpot}). The 
$V_{\rm MDQD}(x,y)$ employed in all our calculations in this paper is displayed in Fig.\ 
\ref{potsdd}(b).

The effective many-body Hamiltonian \cite{macd18,ange21,fu20,yann23} associated with the isolated 
MDQD is given by
\begin{align}
H_{\rm MB} = \sum_{i=1}^N \left\{ \frac{{\bf p}_i^2}{2 m^*} +V_{\rm MDQD}(\br_i) \right\} +
\sum_{i<j}^N \frac{e^2}{\kappa |\br_i-\br_j|},
\label{mbh}
\end{align}
where $m^*$ is the effective mass of the holes and $\kappa$ is the dielectric constant.
A brief outline of the FCI and sS-UHF methodologies, used to solve the corresponding many-body 
Schr\"{o}dinger equation, is presented in Appendices \ref{a1}, \ref{a2}, and \ref{a3}.

{\it FCI results for $N=4$ holes in the double-dot confinements of Fig.\ \ref{potsdd}.\/}
The ground-state FCI charge densities for the four holes confined in the double-MQD of Fig.\ 
\ref{potsdd}(a) and Fig.\ \ref{potsdd}(b) (that is, corresponding to $\nu=2$ filling of the moir\'e 
superlattice), are displayed in Fig.\ \ref{fcicd}(a) and Fig.\ \ref{fcicd}(b), respectively;
the corresponding FCI total spin is found to be $S=0$ with spin projection $S_z=0$.  

Unlike the CDs in Fig.\ \ref{fcicd}(a), which are rather 
ellipsoidal-like, the CDs in each MQD in Fig.\ \ref{fcicd}(b) do exhibit a trilobal deformation, 
reflecting the trilobal deformation of the confining potential [Fig.\ \ref{potsdd}(b)]. More 
importantly, in spite of the Coulombic interaction between the left and right MQD, which is fully 
taken into account via our FCI calculation, the $N=2$ CDs in each MQD of Fig.\ \ref{fcicd}, in both
panels (a) and (b), retain the ring-like shape [albeit pear-like distorted in (b)] found for an 
$N=2$-hole single MQD in our earlier study \cite{yann23}; see also the inset in Fig.\ \ref{pots}(b) 
shown above.   

Naively, these ring-shaped CDs are incompatible with the dumbbell shape of the bonding charge 
distribution of a 'generic' natural molecule (e.g., H$_2$). However, the case here pertains to a 
genuinely quantum-mechanical effect: namely, the 2-hole antipodal arrangement is hidden (unseen) in 
the CDs, but its presence is revealed via the conditional probability distributions (CPDs) (which 
are second-order, density-density correlation functions \footnote{Second-order correlations have 
been used to decipher counterintuitive and spectacular quantum-mechanical behaviors in several 
fields of physics, e.g., the Hong-Ou-Mandel effect in optics \cite{hong87,mand99} or the far-field 
particle coincidence maps in ultracold atoms \cite{yann19}}). Indeed the CPD analysis 
\cite{yann23} of such WMs in MQDs (for $2 \leq N \leq 6$) applied to results obtained
via FCI (exact) calculations, is supplemented and complemented in this paper through the comparative
investigation of exact and mean-field sS-UHF results, with the latter corresponding to approximate 
solutions of the confined quantum few-body problem. Such UHF vs FCI comparative analysis is part of 
a constructive hierarchical approach to the complex few-body problem (see Fig.\ 1 in Ref.\ 
\cite{yann07}).

{\it UHF charge densities for $N=4$ holes in the double-dot confinement of Fig.\ \ref{potsdd}(b).\/} 
Charge densities for the two lowest-energy sS-UHF isomers with $S_z=0$ 
\footnote{The two UHF isomers have very 
close energies, i.e., 526.05 meV and  526.25 meV, respectively. The FCI ground state has a lower 
energy of 505.54 meV. We note that the UHF does not preserve the total spin.} 
(for $N=4$, considered with the same model parameters as in the FCI 
calculations) in the double MQD confining potential of Fig.\ \ref{potsdd}(b) that includes the 
$\sin(3\theta^\prime_k)$ trilobal contributions referenced to the center of each QD, are displayed 
in Fig.\ \ref{uhfcd}(a) and Fig.\ \ref{uhfcd}(b). Unlike the 
electron charge distribution obtained via the exact-diagonalization (FCI) calculation, the sS-UHF 
CDs in Fig.\ \ref{uhfcd}(a) and Fig.\ \ref{uhfcd}(b) exhibit  prominently a dimer of two well 
localized particles within each QD. The difference between these two sS-UHF CDs pertains to the 
different relative orientations between the axes of the two dimers in the right and left MQDs. 
Namely in Fig.\ \ref{uhfcd}(a) the left and right dimers are perpendicular to the $x$-axis, while in
Fig.\ \ref{uhfcd}(b) the left dimer is oriented parallel to the $x$-axis, with the right dimer 
retaining an orientation perpendicular to the $x$-axis. 

It is clear that the sS-UHF CDs do not agree with the FCI CD in Fig.\ \ref{fcicd}(b). This 
disagreement indicates that, in order to obtain a reliable and satisfactory approximate solution, 
it is imperative that further corrective steps, beyond the mean-field level, need to be taken. 
Indeed, a complete theory
of such correctional steps is known under the umbrella term of restoration of broken symmetries 
\cite{yann07,shei21,ringbook}. The full set of corrections \cite{yann07,shei21,ringbook}, which can
produce better beyond-UHF approximate solutions and results for both the CDs and the total energies
is beyond the scope of this paper. Nevertheless, an immediately recognizable and available 
correction is the restoration of the $x$-parity symmetry of the sS-UHF wave function about the 
$y$-axis, which is visibly broken in the CD of Fig.\ \ref{uhfcd}(b). Such an $x$-parity restoration 
can be implemented as described in the following paragraph. 

Denoting the UHF Slater determinant as $\Psi(x,y)$, its mirror image about the $y$ axis is given by
$\Psi(-x,y)$, and the $x$-parity restored wave function is $\propto \Psi(x,y)+p \Psi(-x,y)$, with $p=\pm 1$.
Then, because the Slater determinant $\Psi(-x,y)$ is in general not orthogonal to $\Psi(x,y)$, the 
expectation value of an operator ${\cal O}$ is given by (here we exhibit the lower-energy case, which
was found for the restored wave function with $p = +1$):
\begin{widetext}
\begin{align}
\frac{\langle \Psi(x,y)| {\cal O}| \Psi(x,y) \rangle + 
\langle \Psi(x,y)| {\cal O}| \Psi(-x,y) \rangle +
\langle \Psi(-x,y)| {\cal O}| \Psi(x,y) \rangle + 
\langle \Psi(-x,y)| {\cal O}| \Psi(-x,y) \rangle}
{\langle \Psi(x,y)| \Psi(x,y) \rangle + \langle \Psi(x,y)| \Psi(-x,y) \rangle +
\langle \Psi(-x,y)| \Psi(x,y) \rangle + \langle \Psi(-x,y)| \Psi(-x,y) \rangle}.
\label{oper}
\end{align} 
\end{widetext}
The operator associated with the charge density is a one-body operator, 
$\sum_i^N \delta(\br - \br_i)$, and the charge density is calculated using Eq.\ (\ref{oper})
and the L\"owdin rules \cite{lowd55,verb91} for calculating matrix elements between Slater 
determinants with non-orthogonal orbitals.  The energy of the symmetry-restored wave function [calculated from Eq.\ (\ref{oper}) with ${\cal O} = H_{\rm MB}$] is lower than the mean-field UHF result, reflecting a gain in correlation energy when going beyond the single-determinant wave function of the UHF
method; $H_{\rm MB}$ is the many-body Hamiltonian.

The resulting CD of this $x$-parity restoration is displayed in Fig.\ \ref{uhfcd}(c), and it 
exhibits an overall shape qualitatively similar to the FCI CD in Fig.\ \ref{fcicd}(b) 
\footnote{In spite of this convergence in the shape of CDs, the simple two-Slater-determinant
wave function, $\Psi(x,y)+\Psi(-x,y)$, is still only an approximation to the FCI one. Indeed, the
lowering in total energy is only $\approx 0.012$ meV compared to a lowering of $\approx 21$ meV
that is required to reach the FCI value; see Ref.\ \cite{Note2}.
This remaining discrepancy can be overcome by restoring 
the total spin and by accounting for additional quantum fluctuations through mixing constrained 
Hartree-Fock solutions associated with several different orientations (beyond those parallel or 
perpendicular to the $x$-axis) of the antipodal dimers; this mixing can be carried out by 
following the Griffin-Hill-Wheeler generator-coordinate method 
\cite{hill53,grif57,yann07,shei21,ringbook}, which is even broader in scope than the 
symmetry-restoration methodology.}
This result provides a vivid illustration of the limitation of the sS-UHF method to yield a proper 
description of the Wigner molecules formed in assembled neighboring quantum dots, and the imperative 
need for improvements, such as the one shown here, gained through the application of the 
beyond-mean-field symmetry-restoration corrective step to the sS-UHF solutions. 

\begin{figure}[t]
\centering\includegraphics[width=8.0cm]{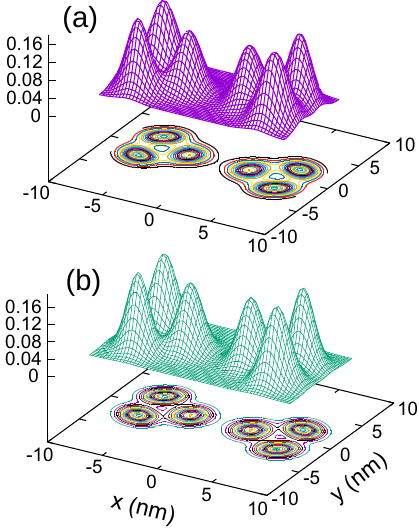}
\caption{
Charge densities for $N=6$ fully polarized holes associated with the corresponding lowest-energy 
state. (a) FCI result (with total spin $S=3$ and spin projection $S_z=3$). 
(b) sS-UHF result (with broken total-spin symmetry and total spin projection $S_z=3$). 
The employed MDQD double-well confinement is displayed in Fig.\ 
\ref{potsdd}(b) (which includes the trilobal deformation within each MQD). In both cases, the CDs 
consist of strongly-pinned (0,3) WMs within each potential well. 
Effective mass $m^*=0.90m_e$ and dielectric constant $\kappa=5$. CDs in units of 1/nm$^2$.
}
\label{fciuhfcd}
\end{figure}

{\it FCI and sS-UHF results for $N=6$ holes in the double-dot confinement of Fig.\ \ref{potsdd}(b).\/}
As uncovered in our earlier study \cite{yann23} on single MQDs, in the case of $\tilde{n}=3$ holes, 
the coincidence of the 3-fold symmetries associated with the $C_3$ intrinsic geometry of the trimer 
WM and with the trilobal crystal-field MQD potential [see Fig.\ \ref{pots}(b)], results in a 
{\it pinned\/}, empty center, three-hump (0, 3) charge density (see Fig.\ 2(b) in Ref.\ 
\cite{yann23}). For the double-dot confinement, the calculated exact FCI [see Fig.\ 
\ref{fciuhfcd}(a)] and approximate sS-UHF [see Fig.\ \ref{fciuhfcd}(b)] charge densities, obtained 
for $N=6$ holes ($\tilde{n}=3$ holes per well, corresponding to $\nu=3$ filling of the moir\'e
superlattice) are qualitatively very similar. Indeed they maintain 
close resemblance to the above-noted pinned 3-fold symmetric configuration in a single MQD. 
This behavior contrasts with that for $N=4$ holes in the same MDQD confinement,
where the FCI CDs in each well exhibit a sliding WM [see Fig.\ \ref{fcicd}(b)] which differ 
drastically from the pinned-WM CDs of the sS-UHF approach [see Figs.\ \ref{uhfcd}(a) and 
\ref{uhfcd}(b)]

{\it Conclusions.\/} The bottom-up research strategy followed in this paper enables a reliable
determination of the influence of interdot Coulomb effects on the formation of quantum WMs in
MQDs associated with integer-filling, $\nu > 1$, supercrystals in moir\'e TMD superlattices. 
Specifically, for the $\nu=2$ case, we demonstrated explicitly that, in spite of the interdot 
Coulombic interaction, the eaxct FCI CDs within each MQD retain the ring-like shape characteristic 
(for a wide range of relevant parameters \cite{yann23}) of a fully isolated MQD. This persisting
behavior, which is deeply counterintuitive and quantum mechanical, is associated with the formation 
of a sliding WM (referred to also as rotating when the confinement exhibits perfect circular
symmetry \cite{yann07}). We also demonstrated that using the mean-field UHF in order to 
account for the interdot Coulomb interaction is an unreliable approach, with the corresponding CDs
portraying orientationally pinned and well localized dumbbell dimers, in contrast to the exact 
result. Notably, we illustrated that the gap between exact and UHF results can be bridged by going 
beyond the mean-field step within a hierarchical strategy that employs the theory of restoration of 
broken symmetries and its generalizations \cite{yann07,shei21}. In contrast to the mean-field 
results, our corrected sS-UHF methodology, with the parity of the ground state wave function being 
restored, yielded (for WMs formed in the coupled MQDs studied here at $\nu=2$) charge densities that 
agree with those obtained via exact (FCI) calculations for that system. 

We conjecture (to be
confirmed both computationally and experimentally) that our conclusions for the $\nu=2$ 
superlattice case would extend to other cases, e.g., to all cases with $4 \leq \nu \leq 6$, where 
our previous study \cite{yann23} determined that a fully quantum mechanical sliding WM (exhibiting a 
ring-like CD) is formed in an isolated MQD instead of an azimuthally pinned WM. Finally, we showed that
the case of $\nu=3$ is an exception to the above behavior due to the commensurability between the 
classical equilibrium configuration of the confined charges and the trilobal $C_3$ crystal-field-type 
anisotropy in each MQD potential pocket.

{\bf NOTE ADDED:}
A recent preprint \cite{crom23} presents measured STM images for 
$\nu=2-4$ integral fillings of hole doped moir\'e TMD
superlattices that are in remarkable agreement with our predictions here (as well as in our Ref.\ 
\cite{yann23}). Ref.\ \cite{crom23} presents also sS-UHF calculations for the WM superlattice and 
comments on their limitations.
  
This work has been supported by a grant from the Air Force Office of Scientific Research (AFOSR) 
under Grant No. FA9550-21-1-0198. Calculations were carried out at the GATECH Center for 
Computational Materials Science. 

\appendix

\section{THE CONFIGURATION INTERACTION METHOD}
\label{a1}

The full configuration interaction (FCI) methodology has a long history, starting in quantum 
chemistry; see Refs.\ \cite{shav98,szabo}. The method was adapted to two dimensional problems and 
found extensive applications in the fields of semiconductor quantum dots
\cite{yann03,szaf03,ront06,yann07.2,yann07,yann09,yann22.2,yann22.3} and of the fractional quantum 
Hall effect \cite{yann04,yann21}.

Our 2D FCI is described in our earlier publications. The reader will find a comprehensive exposition
in Appendix B of Ref.\ \cite{yann22.2}, where the method was applied to GaAs double-quantum-dot 
quantum computer qubits. We specify that, in the application to moir\'{e} DQDs, we keep similar
space orbitals, $\varphi_j(x,y)$, $j=1,2,\ldots,K$, that are employed in the building of the 
single-particle basis of spin-orbitals used to construct the Slater determinants $\Psi_I$, which 
span the many-body Hilbert space [see Eq.\ (B4) in Ref.\ \cite{yann22.2}; the index $I$ counts the 
Slater determinants]. Accordingly, for a moir\'{e} DQD, the orbitals $\varphi_j(x,y)$ are determined
as solutions (in Cartesian coordinates) of the auxiliary Hamiltonian
\begin{equation}
H_{\rm aux}=\frac{{\bf p}^2}{2m^*} + \frac{1}{2} m^* \omega_y^2 y^2
      + \frac{1}{2} m^* \omega_{xk}^2 x_k^{\prime 2},
\label{haux}
\end{equation}
where the index $k=1$ for $x<0$ (left well) and $k=2$ for $x \geq 0$ (right well). 

Following Ref.\ \cite{yann22.2}, we use a sparse-matrix eigensolver based on Implicitly Restarted 
Arnoldi methods to diagonalize the many-body Hamiltonian in Eq.\ (6) of the main text. 

The smooth-neck (one-body) and Coulomb (two-body) matrix elements required for the sparse-matrix 
diagonalization are calculated numerically as described in Ref.\ \cite{yann22.2}. Similarly, the 
matrix elements between the orbitals $\varphi_i(x,y)$ and $\varphi_j(x,y)$ of the trilobal 
(one-body) term in the moir\'{e} DQD confinement [second term in Eq.\ (5) of the main text] are also
calculated numerically.

\section{THE SPIN-AND-SPACE UNRESTRICTED HARTREE-FOCK AND SYMMETRY RESTORATION}
\label{a2}

Early on in the context of 2D materials, the spin-and-space unrestricted Hartree-Fock (sS-UHF) was 
employed in Ref.\ \cite{yann99} to describe formation of Wigner molecules at the mean-field level. 
This methodology employs the Pople-Nesbet equations \cite{szabo,yann07}. The sS-UHF WMs are
self-consistent solutions of the Pople-Nesbet equations that are obtained by relaxing both the 
total-spin and space symmetry requirements. For a detailed description of the Pople-Nesbet equations
in the context of three-dimensional natural atoms and molecules, see Ch.\ 3.8 in Ref.\ \cite{szabo}.
For a detailed description of the Pople-Nesbet equations in the context of two-dimensional
artificial atoms and semiconductor quantum dots, see Sec.\ 2.1 of Ref.\ \cite{yann07}.
Convergence of the self-consistent iterations was achieved in all cases by mixing the input and
output charge densities at each iteration step. The convergence criterion was set to a difference
of $10^{-12}$ meV between the input and output total UHF energies at the same iteration step.    

We note that the book of Szabo and Ostlund \cite{szabo} does not describe the post-Hatree-Fock theory 
of symmetry restoration. For a detailed description of the theory of symmetry restoration, see Sec.\
2.2 in Ref.\ \cite{yann07}.

\section{CHARGE DENSITIES FROM FCI AND UHF WAVE FUNCTIONS}
\label{a3}

The FCI single-particle density (charge density) is the expectation value of a one-body operator
\begin{equation}
\rho({\bf r}) = \langle \Phi^{\rm FCI}
\vert  \sum_{i=1}^N \delta({\bf r}-{\bf r}_i)
\vert \Phi^{\rm FCI} \rangle,
\label{elden}
\end{equation}
where $\Phi^{\rm FCI}$ denotes the many-body (multi-determinantal) FCI wave function, namely,
\begin{equation}
\Phi^{\rm FCI} ({\bf r}_1, \ldots , {\bf r}_N) =
\sum_I C_I \Psi_I({\bf r}_1, \ldots , {\bf r}_N),
\label{mbwf}
\end{equation}
with $\Psi_I({\bf r)}$ denoting the Slater determinants that span the many-body Hilbert space. 

For the sS-UHF case, one substitutes $\Phi^{\rm FCI}$ in Eq.\ (\ref{elden}) with the 
single-determinant, $\Psi^{\rm UHF}({\bf r})$, solution of the Pople-Nesbet equations. 
$\Psi^{\rm UHF}({\bf r})$ is built out from the UHF spin-orbitals whose space part has the form:
\begin{align}
u^\alpha_i=\sum_{\mu=1}^K {\cal C}^\alpha_{\mu i} \varphi_\mu, \;\;\; i=1,\ldots,K,
\end{align}
and
\begin{align}
u^\beta_i=\sum_{\mu=1}^K {\cal C}^\beta_{\mu i} \varphi_\mu, \;\;\; i=1,\ldots,K,
\end{align}
where the expansion coefficients ${\cal C}^\alpha_{\mu i}$ and ${\cal C}^\beta_{\mu i}$ are solutions
of the Pople-Nesbet equations.

\nocite{*}
\bibliographystyle{apsrev4-2}
\bibliography{mycontrols,moire_doubledot_WM_PRB_arx}

\end{document}